\documentclass[aps,prl,twocolumn,superscriptaddress,showpacs]{revtex4-2}
\usepackage{amsmath, amssymb,wasysym,graphicx}
\usepackage{appendix}
\usepackage[ruled]{algorithm2e}

\usepackage[utf8]{inputenc}
\usepackage[T1]{fontenc}
\usepackage{xcolor}

\usepackage{diagbox}
\usepackage{tabularx}

\IfFileExists{newtxtext.sty}
{\usepackage{newtxtext,newtxmath}}
{\IfFileExists{stix.sty}
	{\usepackage{stix}}
	{\IfFileExists{mathptmx.sty}
		{\usepackage{mathptmx}}{} } }

\usepackage{textcomp}

\usepackage{bm}

\usepackage{booktabs,siunitx}

\usepackage{color}
\definecolor{LinkColor}{rgb}{0.256,0.439,0.588}
\usepackage{hyperref}
\hypersetup{
	pdfauthor={good guys},
	pdftitle={good title},
	colorlinks=true,
	citecolor=LinkColor,
	linkcolor=LinkColor,
	urlcolor=LinkColor
}

\newcommand{\bra}[1]{\langle#1\rvert}
\newcommand{\ket}[1]{\lvert#1\rangle}

\def\bkp{{ \mathbf{k}^{\prime} }}

\def\bk{{\mathbf{k}}}

\def\bK{{\mathbf{K}}}

\def\bq{{\mathbf{q}}}

\def\bG{{\mathbf{G}}}

\usepackage{multirow}

\begin{document}
\title{Thermodynamic characteristic for correlated flat-band system with quantum anomalous Hall ground state}

\author{Gaopei Pan}
\affiliation{Beijing National Laboratory for Condensed Matter Physics and Institute of Physics, Chinese Academy of Sciences, Beijing 100190, China
}
\affiliation{School of Physical Sciences, University of Chinese Academy of Sciences, Beijing 100049, China}

\author{Hongyu Lu}
\affiliation{Department of Physics and HKU-UCAS Joint Institute
	of Theoretical and Computational Physics, The University of Hong Kong,
	Pokfulam Road, Hong Kong SAR, China}

\author{Heqiu Li}
\affiliation{Department of Physics, University of Toronto, Toronto, Ontario M5S 1A7, Canada}

\author{Xu Zhang}
\affiliation{Department of Physics and HKU-UCAS Joint Institute
	of Theoretical and Computational Physics, The University of Hong Kong,
	Pokfulam Road, Hong Kong SAR, China}

\author{Bin-Bin Chen}
\affiliation{Department of Physics and HKU-UCAS Joint Institute
	of Theoretical and Computational Physics, The University of Hong Kong,
	Pokfulam Road, Hong Kong SAR, China}

\author{Kai Sun}
\email{sunkai@umich.edu}
\affiliation{Department of Physics, University of Michigan, Ann Arbor, Michigan 48109, USA}

\author{Zi Yang Meng}
\email{zymeng@hku.hk}
\affiliation{Department of Physics and HKU-UCAS Joint Institute
	of Theoretical and Computational Physics, The University of Hong Kong,
	Pokfulam Road, Hong Kong SAR, China}

\begin{abstract}
While the ground state phase diagram of the correlated flat-band systems have been intensively investigated, the dynamic and thermodynamic properties of such lattice models are less explored, but it is the latter which is most relevant to the experimental probes (transport, quantum capacitance and spectroscopy) of the quantum moir\'e materials such as twisted bilayer graphene and transition metal dichalcogenides. Here we show, by means of momentum-space quantum Monte Carlo and exact diagonalization, there exists a unique thermodynamic characteristic for the correlated flat-band models with interaction-driven quantum anomalous Hall (QAH) ground state, namely, the transition from the QAH insulator to the metallic state takes place at a much lower temperature compared with the zero-temperature single-particle gap generated by the long-range Coulomb interaction. Such low transition temperature comes from the proliferation of excitonic particle-hole excitations, which ``quantum teleport'' the electrons across the gap between different topological bands to restore the broken time-reversal symmetry and give rise to a pronounced enhancement in the charge compressibility. Future experiments, to verify such generic thermodynamic characteristics, are proposed.
\end{abstract}
\date{\today}
\maketitle

{\it Introduction}\,---\,  Quantum moir\'e systems, bestowed with the quantum geometry of wavefunctions -- manifested in the distribution of Berry curvature in the flat bands -- and strong long-range Coulomb electron interactions, exhibit a rich quantum phase diagram including correlated insulating, unconventional metallic and superconducting phases, thanks to the high tunability by twisting angles, gating and tailored design of the dielectric environment \cite{tramblyLocalization2010,tramblyNumerical2012,bistritzerMoire2011,Santos2012,lopesGraphene2007,caoUnconventional2018,shenCorrelated2020,xieSpectroscopic2019,KhalafCharged2021,KevinStrongly2020,pierceUnconventional2021,caoCorrelated2018,liaoValence2019,liaoCorrelated2021,luSuperconductors2019,moriyamaObservation2019,chenTunable2020,rozhkovElectronic2016,ChatterjeeSkyrmion2020,kerelskyMaximized2019,rozenEntropic2021,tomarkenElectronic2019,soejimaEfficient2020,liuSpectroscopy2021,KhalafSoftmodes2020,ZondinerCascade2020,saitoPomeranchuk2021,GhiottoCriticality2021,SchindlerTrion2022,WangTMD2020,Parkchern2021,liaoCorrelation2021,anInteraction2020,huangGiant2020,liLattice2021}. In addition to this complex ground-state phase diagram with possibly different pairing mechanism and symmetry breaking patterns~\cite{zhangSuperconductivity2021,KhalafCharged2021,wagnerGlobal2022,zhangCorrelated2022,liuTheories2021}, recent theoretical studies~\cite{liaoCorrelated2021,liaoCorrelation2021,chenRealization2021,linExciton2022,panSport2022} indicates that
correlated flat bands also exhibit unique  thermodynamic and quantum dynamic responses, fundamentally different from conventional correlated electron lattice model systems, such as the Hubbard-type model.

Previous results~\cite{chenRealization2021} based on a real-space effective model~\cite{kangStrong2019,koshino2018,poOrigin2018} for twisted bilayer graphene (TBG) at 3/4 filling, via density matrix renormalization group computation, successfully identify a quantum anomalous Hall (QAH) state as the long-sought-after topological Mott insulator (TMI)~\cite{raghu2008,jiaEffect2013,capponiPhase2015}, as the ground state of the interaction-only system spontaneously breaks the time-reversal symmetry and acquires a finite Chern number. Then by means of thermal tensor network and the perturbative field-theoretical approaches~\cite{linExciton2022}, the finite-$T$ phase diagram and the dynamical properties of the real-space TBG model have been revealed, which contains the QAH and charge density wave insulators at low-$T$, and an Ising transition separating them from the high-$T$ symmetric phases. Because of the proliferation of excitons -- particle-hole bound states -- this phase transition takes place at a significantly reduced temperature (at the scale of a few meV) than the mean-field estimation of the topological band gap (at the scale of a few tens of meV). Between these two energy scales, an exciton-proliferated phase is observed, which acquires distinctive experimental signatures in charge compressibility and optical conductivities close to the transition~\cite{linExciton2022}. 

Although the aforementioned real-space effective lattice models offers a clear physics picture and important insights about moir\'e systems, certain limitations in these models, such as the absence of the quantum metric of the quantum wavefunction and the truncation of the long-range Coulomb interaction, makes it a nontrivial task to directly connect these model calculations with realistic quantum moir\'e materials. For this purpose, unbiased computations of the  Bistritzer-MacDonald (BM) continuum model~\cite{tramblyLocalization2010,bistritzerMoire2011,tramblyNumerical2012}, originated from the flat-bands in momentum space and subjected to the truely long-range Coulomb interactions, are highly desirable to explore the thermodynamic characteristic of the correlated flat-band systems and provide guidance to future experiments.

In this work, we employ the momentum-space quantum Monte Carlo (QMC) method~\cite{zhangMomentum2021,panDynamical2022,hofmannFermionic2022}, supplemented with exact diagonalization (ED)~\cite{liSpontaneous2021}, to systematically investigate the thermodynamic properties of realistic correlated flat-band TBG model at 3/4 filling. We find that in analogy its real-space cousin, the continuous model also exhibit two distinct temperature scales, defined by the transition temperature ($T_c \sim 4$ meV) and topological gap ($\Delta \sim 17$ meV).
Although the $T=0$ ground state is a topologically-nontrivial QAH-TMI phase with a large gap $\sim 17$ meV, this topological phase melts at only $\sim 4$ meV, through a continuous  phase transition that belongs to the Ising universality class. At intermediate temperature $T_c < T < \Delta$,  excitonic collective excitations teleport valence electrons across the topological band gap to the opposite Chern bands and restore the broken time-reversal symmetry. Such process generates pronounced enhancement in the charge compressibility, readily detectable via the quantum capacitance measurements~\cite{tomarkenElectronic2019,peterThe2020}. The associated temperature dependence of the electronic spectral function $A(\mathbf{k},\omega)$ are also obtained with scrutiny. Our work therefore confirms the unique thermodynamic characteristic of the correlated flat-band systems and offer the mechanism for the smearing of the many-electron state topology by proliferating collective exciton excitations and open an avenue for controlled many-body computation on finite-temperature states in quantum moiré systems.

\begin{figure}
\includegraphics[width=\linewidth]{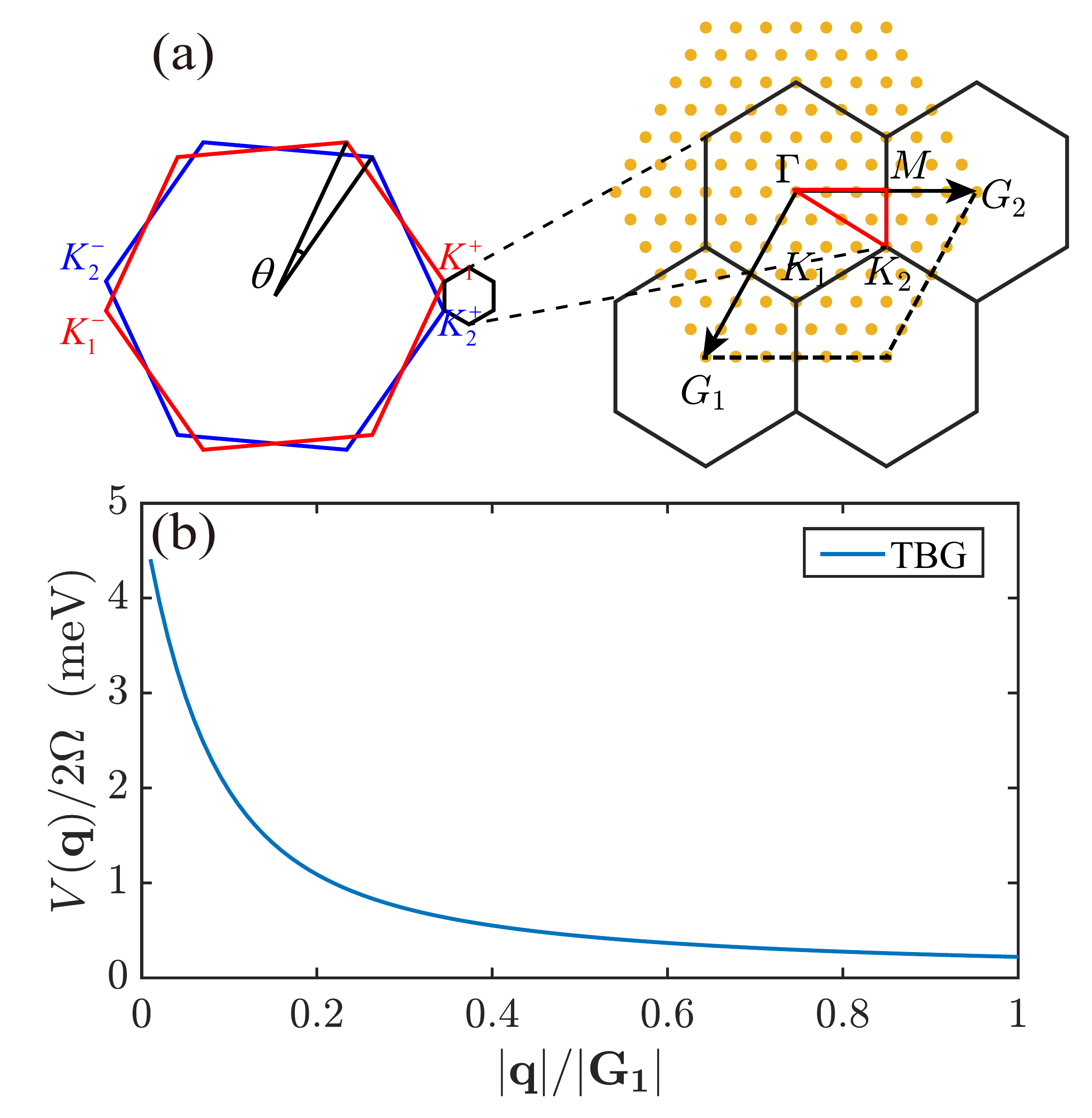}
\caption{(a) Schematic TBG setting in momentum space and its moir\'e Brilliouin zone (mBZ) at one valley. The red solid line marks the high-symmetry path $\Gamma-M-K_1(K_2)-\Gamma$. $\bG_1$ and $\bG_2$ are the reciprocal lattice vectors of the mBZ. Yellow dots mark possible momentum transfer in QMC simulations, $\bq+\bG$. Because the form factor decays exponentially with $\bG$~\cite{bernevig2020tbg5}, scatterings with momentum transfer larger than this cut-off are ignored. Here we show a $6\times 6$ mesh in the mBZ, with 126 allowed momentum transfers. (b) The decay of the single gate Coulomb interaction for the systems with the input parameters in the main text.}   
\label{fig:fig1}
\end{figure}

\begin{figure}[!h]
\centering
\includegraphics[width=\linewidth]{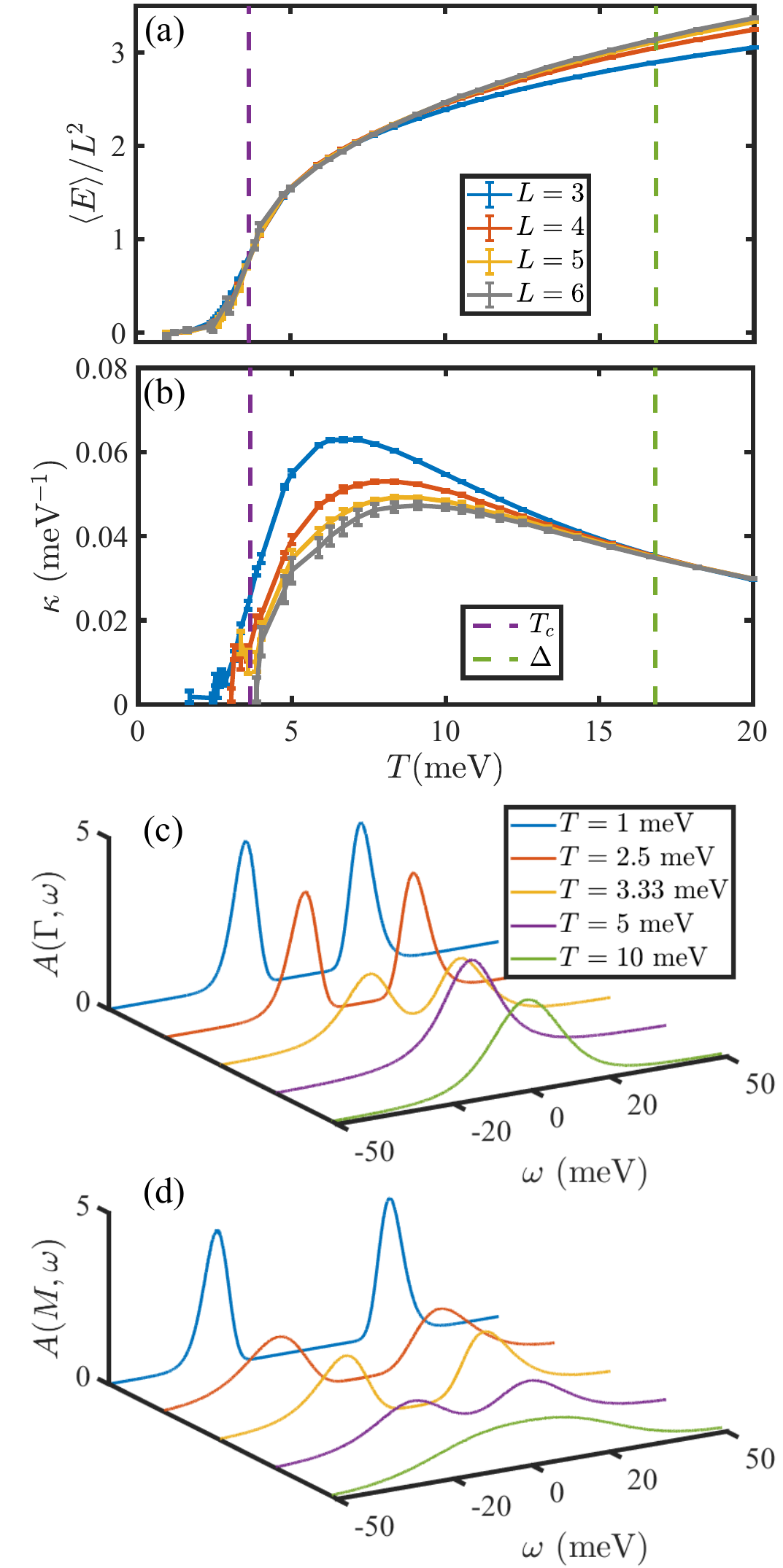}
\caption{(a) and (b) The energy density and compressibility $\kappa$ of TBG model in Eq.~\eqref{eq:eq2} with $L=3,4,5,6$ as a function of temperature. The single-particle gap $\Delta$ is denoted as the green dashed line. The QAH transition temperature $T_c$ (determined in Fig.~\ref{fig:fig3} below) is denoted by the purple dashed line. (c) and (d) Single-particle spectral function $A(\mathbf{k},\omega)$ obtained form QMC-SAC at $\mathbf{k}=\Gamma$ and $M$ with $L=4$. Below the temperature scale of the single-particle gap, the system remains gapless due to the proliferation of excitons. The spectral gap only arises below the thermal transition temperature, when the system enters the QAH-TMI phase.}
	\label{fig:fig2}
\end{figure}

\begin{figure*}[htp!]
\centering
\includegraphics[width=\linewidth]{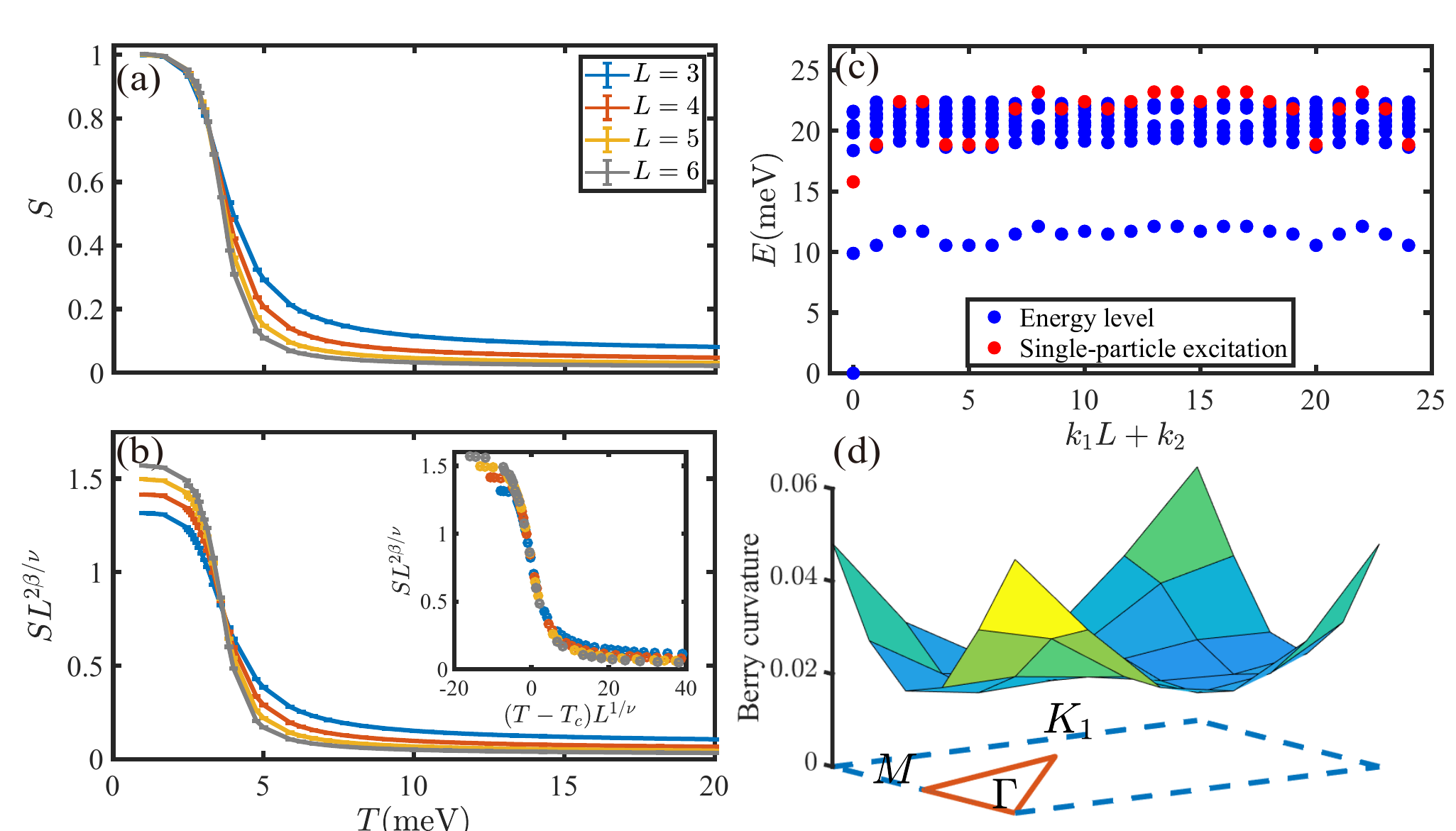}
\caption{(a) and (b) Correlation fucntion of the QAH order parameter $S$ versus temperature for $L=3,4,5,6$ from QMC. The finite-size scaling using 2D Ising exponents determines the thermal transition temperature at $T_c=3.65(5)$ meV. Inset of (b) shows a perfect data collapse. (c) Energy spectrum of TBG system from exact diagonalization. The blue dots represent the energy of each many-body state with $N_k=25$ in $5\times 5$ mesh of mBZ, which include the ground state at zero energy, the low-lying exciton states ($\sim 10$ meV) and excitations with higher energy. The red dots represent single particle excitation, which is at much higher energy than that of the collective excitons. (d) Berry curvature from QMC for $L=6$ at $T=1.25$ meV. Chern number $C=\sum_{\square} F_{\square}=1$.}
\label{fig:fig3}
\end{figure*}

{\it Model and Method}\,---\,
We construct the continuum model for the magic angle TBG with flat-bands and long-range Coulomb interaction, following the convention used in Refs.~\cite{song2020tbg2,bernevig2020tbg3,tramblyLocalization2010,bistritzerMoire2011,tramblyNumerical2012,rozhkovElectronic2016,lopesGraphene2007,Santos2012,YiZhang2020}. The Hamiltonian of BM model in the plane wave basis is
\begin{equation}
	\begin{aligned}
		&H_{B M,\mathbf{k},\mathbf{k {}^\prime}}=\delta_{\mathbf{k},\mathbf{k {}^\prime}}\left(\begin{array}{cc} 
			-\hbar v_F ({\bk}-\bK_1^{+}) \cdot \pmb{\sigma}  &  U_0  \\
			U_0^\dagger  & -\hbar v_F ({\bk}-\bK_2^{+}) \cdot \pmb{\sigma}
		\end{array}\right) \\ 
		+&\left(\begin{array}{cc} 
			0  &  U_1 \delta_{\mathbf{k},\mathbf{k {}^\prime}- \bG_1 }+U_2 \delta_{\mathbf{k},\mathbf{k {}^\prime}-(\bG_1+\bG_2) }  \\ 
			U_1^{ \dagger} \delta_{\mathbf{k},\mathbf{k {}^\prime}+ \bG_1 }+U_2^{ \dagger} \delta_{\mathbf{k},\mathbf{k {}^\prime}+(\bG_1+\bG_2) }  & 0
		\end{array}\right)
	\end{aligned}
	\label{eq:eq1}
\end{equation}
Here we only consider one valley and one spin, where the Pauli matrices $\boldsymbol{\sigma}=\left( \sigma_{x}, \sigma_{y}\right)$ are from the A, B sublattices of the monolayer graphene. $\mathbf{G}_{1}$ and $\mathbf{G}_{2}$ are the reciprocal vectors of moiré Brillouin zone (mBZ) as shown in Fig.~\ref{fig:fig1} (a). $\mathbf{K}_{1}^{\tau}$ and $\mathbf{K}_{2}^{\tau}$ are the Dirac points of the bottom and top graphene layers respectively, which are twisted by the angles $\mp \frac{\theta}{2}$. The interlayer tunneling between the Dirac states is described by the matrices $U_{0}=\left(\begin{array}{ll} u_0 & u_{1} \\ u_{1} &  u_0\end{array}\right)$, $U_{1}=\left(\begin{array}{cc} u_0 & u_{1} e^{- \frac{2 \pi}{3} i} \\ u_{1} e^{ \frac{2 \pi}{3} i} &  u_0\end{array}\right)$ and $U_{2}=\left(\begin{array}{cc} u_0 & u_{1} e^{ \frac{2 \pi}{3} i} \\ u_{1} e^{- \frac{2 \pi}{3} i} &  u_0\end{array}\right)$ where $u_0$ and $u_{1}$ are the intra- and  inter-sublattice interlayer tunneling amplitudes. The flatness of the lowest two bands in the chiral limit $\left(u_{0}=0\right)$ is determined by the dimensionless parameter $\alpha=\frac{u_{1}}{\hbar v_{F} k_{\theta}}$ with small $\theta$ approximation $k_{\theta}=4 \pi \theta  /\left(3 a_{0}\right)$ and the lattice constant of the monolayer graphene $a_{0}=0.246 \mathrm{~nm}$. In this paper, we choose $\hbar v_{F} / a_{0}=2.37745 \mathrm{eV}$, the twist angle $\theta=1.08^{\circ}$ and $u_{1}=0.11 \mathrm{eV}$ which leads to the first magic angle where the lowest two bands become completely flat in the chiral limit.

Once projecting the long-range Coulomb interaction onto the flat bands, the Hamiltonian becomes,
\begin{equation}
	H=\frac{1}{2 \Omega} \sum_{\mathbf{q}, \mathbf{G},|\mathbf{q}+\mathbf{G}| \neq 0} V(\mathbf{q}+\mathbf{G}) \delta \rho_{\mathbf{q}+\mathbf{G}} \delta \rho_{-\mathbf{q}-\mathbf{G}}
	\label{eq:eq2}
\end{equation}
where $V(\mathbf{q})=$ $\frac{e^{2}}{4 \pi \varepsilon} \int d^{2} \mathbf{r}\left(\frac{1}{\mathbf{r}}-\frac{1}{\sqrt{\mathbf{r}^{2}+d^{2}}}\right) e^{i \mathbf{q} \cdot \mathbf{r}}=\frac{e^{2}}{2 \varepsilon} \frac{1}{q}\left(1-e^{-q d}\right)$ is the long-ranged single gate Coulomb interaction  with $d/2$ the distance between graphene layer and single gate~\cite{liuNematic2021} and $\epsilon =7 \varepsilon_{0}$ is the dielectric constant. And $\delta \rho_{\bq+\bG}=\sum_{\bk, m_{1}, m_{2}} \lambda_{m_1,m_2}(\bk,\bk+\bq+\bG) \left(d_{\bk, m_{1}}^{\dagger} d_{\bk+\bq, m_{2}}  -\frac{1}{2} \delta_{q,0}\delta_{m_1,m_2}\right)$. $m_1, m_2=1,2$ are the band indices. $d_{\boldsymbol{k}, m}^{\dagger}$ is the creation operator for a Bloch eigenstate, $\left|u_{\boldsymbol{k}, m}\right\rangle$, which is the eigenstate of the BM Hamilton in Eq.~\eqref{eq:eq1}. The form factor is defined as $\lambda_{m_{1}, m_{2}}(\mathbf{k}, \mathbf{k}+\mathbf{q}+\mathbf{G}) \equiv\left\langle u_{\boldsymbol{k}, m_1} \mid u_{\boldsymbol{k}+\bq+\bG, m_2}\right\rangle$. The moiré lattice vector $L_{M}=\frac{a_{0}}{2 \sin \left(\frac{\theta}{2}\right)}$, the mBZ reciprocal lattice vector $|\mathbf{G}_0|=\frac{4 \pi}{\sqrt{3} L_{M}}$ and the area of the mBM is $\Omega=N_k \frac{\sqrt{3}}{2} L_{M}^{2}$ with $N_k=L\times L$ the number of $\mathbf{k}$ points in $\mathrm{mBZ}$, for example, $N_k=36$ for $6\times 6$ meshs in Fig.~\ref{fig:fig1} (a). We use distance $d=40$ nm and the typical energy scale of $V(|\mathbf{q}|)/2\Omega$ for $L=6$ is shown in Fig.~\ref{fig:fig1}(b).

To compute the dynamic and thermodynamic properties of the TBG model in Eq.~\eqref{eq:eq2}, we employ the recently developed momentum space QMC method~\cite{zhangMomentum2021,hofmannFermionic2022}, which can fully incorporate the flat-band topological wavefunction and the long-range Coulomb interaction and our {\it sign bound theory} can  prove there is at most polynomial sign problem for QMC at integer fillings~\cite{zhangSign2021,panSign2022}, rendering the computational complexity also polynomial. The QMC solves the finite size systems in a path-integral of partition function in an unbiased manner and we have simulated $L=3,4,5,6$ and $T\in [0.5,20]$ meV systems. In the QMC simulation, we compute the charge compressibility $\kappa$ as a function of temperature $T$ as $\kappa=\frac{1}{n_{0} N_k} \frac{d N}{d \mu}=$ $\beta \frac{\left\langle N^{2}\right\rangle-\left\langle N\right\rangle^{2}}{n_{0} N_k}$, where $n_{0}$ is the particle density, which is 1/2 at low-temperature 
and  $N=\sum_{\mathbf{k}_{1}, m_{1}} d_{\mathbf{k}_{1}, m_{1} }^{\dagger} d_{\mathbf{k}_{1}, m_{1}} $ is the particle number operator. We also compute the temperature dependence of the spetral functions from the stochastic analytic continuation (SAC) of the imaginary time Green's function from QMC~\cite{panDynamical2022,yanRelating2021,zhangSuperconductivity2021,zhouAmplitude2021,yanTopological2021} and perform the finite size scaling of the order parameter for the QAH phase and find a good agreement with Ising universality. The topological aspect of the QAH phase can be also seen from the computed Berry curvature below the transition temperature, and from which the Chern numbe can then be calculated as $C=\sum_{\square} F_{\square}$, where $
F_{\square}$
is the Berry curvature in the finite size mBZ obtained from the Green's function in the QMC simulations~\cite{wangSimplified2012,fukuiChern2005}. The detailed QMC implementation of the TBG model and the analyses of different physical observables are given in the Supplemental Material(SM)~\cite{suppl}.

In the chiral limit the two flat moir\'e bands can be combined to obtain a Chern band basis~\cite{Ledwith2020,song2020tbg2,LiTBCB2021} such that each band has Chern number $1$ or $-1$. More importantly, the wave function overlap between these two bands is exactly zero, indicating the form factor in the Chern basis is diagonal in the band index $\lambda_{mn}(\mathbf k,\mathbf k')\sim\delta_{mn}$. The diagonal form factor gives rise to an emergent conservation law: the conservation of the Chern polarization $P_C$ defined by $P_C=N_+-N_-$,  where $N_\pm$ is the number of electrons in the positive and negative Chern bands respectively and $N_++N_-=N$. We note $P_C/N_k$ is the order parameter for the QAH phase and we compute its correlation functions in the QMC simulation to determine the Ising transition temperature precisely (see Fig.~\ref{fig:fig3}). 

We have also incorporated the exact diagonalization (ED) for the same models, where the single-particle and the exciton gaps for different cluster sizes are computed.


{\it Results}\,---\, Fig.~\ref{fig:fig2} (a) and (b) show the energy densities and charge compressibility $\kappa$ as a function of temperature. Here we denote the zero-temperature single-particle gap $\Delta\sim 17$ meV and the thermal transition temperature $T_c\sim 4$ meV using the green and purple dashed lines respectively. The gap $\Delta$ is obtained from the exact solution of the chiral limit (see SM~\cite{suppl}). The transition temperature $T_c$, from scaling analysis of the QAH order parameter shown below, is found to be dramatically smaller than $\Delta$. This small $T_c$ is consistent with thermodynamic and dynamic measurements, e.g., the maximum slop in Fig.~\ref{fig:fig2} (a), and both the compressibility $\kappa$ and the fermion spectrum function $A(\mathbf{k},\omega)$ at $\mathbf{k}=\Gamma$ and $M$ in Fig.~\ref{fig:fig2} (c-d) all indicate that $T_c$ marks the metal-insulator transition, below which the system becomes incompressible and insulating. 

As mentioned above, in the Chern basis, the QAH order parameter is defined as $P_C/N_k$, and its correlation function is
\begin{equation}
S \equiv \frac{\left\langle \left(N_{+}-N_{-}\right)^2\right\rangle}{N_k^{2}}=\frac{1}{N_k^{2}}\left\langle \left(\sum_{\boldsymbol{k}} d_{\boldsymbol{k},+}^{\dagger} d_{\boldsymbol{k},+}-\sum_{\boldsymbol{k}}d_{\boldsymbol{k},-}^{\dagger} d_{\boldsymbol{k},-}\right)^2\right\rangle.
\end{equation}
In Fig.~\ref{fig:fig3}(a), we plot its the temperature dependence. It can be seen that the corresponding order parameters rise at low temperature, which means that spontaneous time-reversal symmetry breaking occurs, and the systems enters the QAH-TMI phase. The thermal phase belongs to the 2D Ising universality class with critical exponents $\beta = 1/8$ and $\nu = 1$. Close to the critical temperature, $S$ are expected to obey the scaling forms $S(T,L)=L^{-2\beta/\nu} f((T-T_c) L^{1/\nu})$ where $f$ is the scaling function.  In Fig.~\ref{fig:fig3} (b), we rescale the data using $S(T,L)L^{2\beta/\nu}= f((T-T_c) L^{1/\nu})$, and the crossing point of different system sizes $L=3,4,5,6$ gives rise to the transition temperatures $T_c=3.65(5)$ meV, which corresponds to the purple dashed lines in Fig.~\ref{fig:fig2} (a) and (b). The inset of Fig.~\ref{fig:fig3}(b) shows the perfect data collapse of the $S$.

Such a small $T_c$ compared with the interaction-driven single-particle gap at the exact ground state is clearly beyond single-particle physics and indicates the importance of many-body fluctuations. 
To pinpoint the origin of this low $T_c$, we performed ED simulations and find that it is due to excitonic excitations in the particle-hole channel, which occurs in our system at a much lower energy in comparison to single-particle excitations and thus lowered the $T_c$. Fig.~\ref{fig:fig3} (c) presents the ED energy spectra, with the blue dots the energy levels with $N_k=25$ electrons in $5\times 5$ mesh resolved by the total momentum $\mathbf k=k_1\mathbf G_1/L+k_2\mathbf G_2/L$. The spectrum for many-body states with Chern polarization $P_C$ is degenerate with those with $-P_C$, and only states with $P_C>0$ are shown. The ground state has zero energy and $P_C=N$, which is fully Chern-polarized. Low energy excitations denoted by the blue dots around 10 meV has $P_C=N-2$, which consists of $N-1$ electrons in the $N_{+}$ Chern band and one electron in the $N_{-}$ Chern band. These states correspond to exciton states in which one of the electrons is excited ("teleported") to the other Chern band from the fully Chern-polarized ground state. The states with lower $P_C$ appear in higher energy in the spectrum. The red dots represent the single-particle excitation, which is obtained from ED computation for a system of the same size and with $N+1$ electrons, and is consistent with the exact gap size $\Delta \sim 17$ meV. For particle-hole excitations (blue dots), although most of them are at energy close to (or higher than) single-particle excitations, one exciton branch arises at very low energy below the single particle gap. These excitons open up a channel for thermal fluctuations at $T_c<T<\Delta$, which destabilize QAH ground state and reduces the Chern polarization (i.e., destroy the quantized Hall conductivity and recovers the time-reversal symmetry).  Only when $T<T_c$, the topological nature of the system manifests, as can be seen in the QMC results of the Berry curvature for $6\times 6$ system at $T=1.25$ meV in Fig.~\ref{fig:fig3} (d), if one integrate over the mBZ, the obtained $C=1$.

{\it Discussion}\,---\, 
In this study, we show that at $3/4$ filling and $T=0$, a QAH-TMI state emerges in magic angle TBGs, due to the interplay between flat band quantum wavefunctions and Coulomb interactions. In analogy to its real-space-model cousin~\cite{linExciton2022}, this QAH-TMI phase melts at a very low temperature, much lower than the zero-temperature gap $\Delta$. We identify that low-energy excitonic states are the origin of this low transition temperature.
At $T_c<T<\Delta$, these excitonic collective modes assist the valence electron to tunnel across the topological band gap, redistribute among the opposite Chern bands and restore the time-reversal symmetry, such process generates great enhancement in the charge compressibility, readily detectable via the quantum capacitance measurements. This behavior is qualitatively different from the conventional weakly interacting topological band insulator or correlated lattice models, such as the Hubbard-type models with local interactions. Its application and extension to both finite temperature experiments in quantum moir\'e materials and the model computation to fully understand the rich phase diagram away from integer fillings, will be anticipated in the near future.

{\it Acknowledgments}\,---\,
We thank Xiyue Li, Wei Li and Tao Shi in the previous collaborations in real-space TBG model, and Yi Zhang and Jian Kang for discussions on the computation of Chern number in moir\'e systems. GPP, HYL, XZ, BBC and ZYM acknowledge the support from the Research Grants Council of Hong Kong SAR of China (Grant Nos. 17303019, 17301420, 17301721, AoE/P-701/20 and  17309822), the K. C. Wong Education Foundation (Grant No. GJTD-2020-01), and the Seed Funding “Quantum-Inspired explainable-AI” at the HKU-TCL Joint Research Centre for Artificial Intelligence. We thank the Computational Initiative at the Faculty of Science and HPC2021 system under the Information Technology Services at the University of Hong Kong, and the Tianhe-II platform at the National Supercomputer Center in Guangzhou for their technical support and generous allocation of CPU time.
\bibliography{ref}
\bibliographystyle{apsrev4-2}

\newpage

\section{Supplementary Materials}
In this Supplementary Material, we discuss the Berry curvature and the algebraic sign problem for the TBG system in QMC simulations. We find that at low-$T$ the system is inside the interaction-driven QAH-TMI phase, with Chern number $C=1$. The single-particle gaps, obtained from the exact solution as well as QMC Green's functions, are also shown to be at the energy scale of Coulomb interaction, and consequently much higher than the thermal transition temperature where the excitons are proliferated. In addition, we introduce the Chern Basis employed both in QMC and exact diagonalization (ED) calculations, in which the order parameter of QAH can be written in a simpler form. We also present the ED results in the Chern basis, which contains the single-particle and the exciton gaps for different cluster sizes.

\subsection{\uppercase\expandafter{\romannumeral1}. Berry Curvature}
\label{sec:SMSeci}

To verify the QAH-TMI ground state, we use QMC data to calculate the Chern number~\cite{wangSimplified2012,fukuiChern2005}.

We construct a gauge-independent algorithm and the Chern number is $C=\sum_{\square} F_{\square}$, where
\begin{equation}
\begin{aligned}
F_{\square}&=\frac{1}{2 \pi} \mathrm{Im} \ln \left(\left\langle\psi(\boldsymbol{k})| \psi\left(\boldsymbol{k}+\delta \boldsymbol{k}_{1}\right)\right\rangle \right. \\
& \left\langle\psi\left(\boldsymbol{k}+\delta \boldsymbol{k}_{1}\right)| \psi\left(\boldsymbol{k}+\delta \boldsymbol{k}_{1}+\delta \boldsymbol{k}_{2}\right)\right\rangle \\
&\left\langle\psi\left(\boldsymbol{k}+\delta \boldsymbol{k}_{1}+\delta \boldsymbol{k}_{2}\right)| \psi\left(\boldsymbol{k}+\delta \boldsymbol{k}_{2}\right)\right\rangle\\
&\left.\left\langle\psi\left(\boldsymbol{k}+\delta \boldsymbol{k}_{2}\right)| \psi\left(\boldsymbol{k}\right)\right\rangle
\right) 
\end{aligned}
\end{equation}
here $|\psi(\bk)\rangle$ is the normalized wave function of the occupied Bloch band. In the Berry curvature $F_{\square}$, $\delta k_{1}$ and  $\delta k_{2}$ are the momentum footstep in the finite size mesh.

In finite temperature QMC simulation, we consider the projection operator $\hat{P}(\mathbf{k})$ which is constructed
using the eigenvectors $\{|e_{1,2}(\bk) \rangle\}$ at $\bk$ with energies below the Fermi energy:
\begin{equation}
\hat{P}(\mathbf{k})=\sum_{\mu_{m}>0}\left|\phi_{m}(0, \mathbf{k})\right\rangle\left\langle\phi_{m}(0, \mathbf{k})\right|=\sum_{i,j} |e_{i} (\bk) \rangle \, P_{ij}(\bk) \,\langle e_{j} (\bk) |
\end{equation} 
where $P(\bk)$ is the matrix at basis $\{|e_{1,2}(\bk) \rangle\}$ and $\left|\phi_{m}(0, \mathbf{k})\right\rangle$ is the eigrnvector of zero-frequency single-particle Green’s
functions $G\left(i \omega_{0}, \mathbf{k}\right)$: 
\begin{equation}
G\left(i \omega_{0}, \mathbf{k}\right)\left|\phi_{m}(0, \mathbf{k})\right\rangle=\mu_{m}(0, \mathbf{k})\left|\phi_{m}(0, \mathbf{k})\right\rangle.
\end{equation}

In the usual Monte Carlo calculation, we have:
\begin{equation}
\begin{aligned}
F_{\square}&=\frac{1}{2 \pi} \mathrm{Im} \ln \operatorname{Tr}\left[\hat{P}(\boldsymbol{k})  \hat{P}\left(\boldsymbol{k}+\delta \boldsymbol{k}_{1}\right) \hat{P}\left(\boldsymbol{k}+\delta \boldsymbol{k}_{1}+\delta \boldsymbol{k}_{2}\right)  \hat{P}\left(\boldsymbol{k}+\delta \boldsymbol{k}_{2}\right)\right]\\
&=\frac{1}{2 \pi} \mathrm{Im} \ln \operatorname{Tr} \left[P(\boldsymbol{k})  P\left(\boldsymbol{k}+\delta \boldsymbol{k}_{1}\right) P\left(\boldsymbol{k}+\delta \boldsymbol{k}_{1}+\delta \boldsymbol{k}_{2}\right)  P\left(\boldsymbol{k}+\delta \boldsymbol{k}_{2}\right)\right],
\end{aligned}
\end{equation}
however in this paper, we can not multiply matrix elements $P_{i,j}$ due to the different choice of the basis of different momentum points after projection onto the flat-bands. To fix this problem, we introduce a link matrix $V_{i,j}(\bk_1,\bk_2)=\langle e_i(\bk_1) |e_j(\bk_2) \rangle$ and modify the Berry curvature to 
 \begin{equation}
 \begin{aligned}
 F_{\square}=\frac{1}{2 \pi} \mathrm{Im} \ln \operatorname{Tr}\left[ \right.& \left.P(\boldsymbol{k})  V\left(\boldsymbol{k}, \boldsymbol{k}+\delta \boldsymbol{k}_{1}\right)  P\left(\boldsymbol{k}+\delta \boldsymbol{k}_{1}\right) \right.\\& V\left(\boldsymbol{k}+\delta \boldsymbol{k}_{1}, \boldsymbol{k}+\delta \boldsymbol{k}_{1}+\delta \boldsymbol{k}_{2}\right)P\left(\boldsymbol{k}+\delta \boldsymbol{k}_{1}+\delta \boldsymbol{k}_{2}\right) \\
 & V\left(\boldsymbol{k}+\delta \boldsymbol{k}_{1}+\delta \boldsymbol{k}_{2}, \boldsymbol{k}+\delta \boldsymbol{k}_{2}\right)P\left(\boldsymbol{k}+\delta \boldsymbol{k}_{2}\right)\\
 &\left. V\left(\boldsymbol{k}+\delta \boldsymbol{k}_{2}, \boldsymbol{k}\right)\right].
 \end{aligned}
 \end{equation}

This is because projecting onto flat band means that now we are in projecting band basis $\left \{|e_{1,2}(\bk) \rangle \right\} =\sum_{i} U(\bk)_{m_{1,2},i}|i \rangle $ in which the BM Hamiltonian $H_{BM}(\bk)$ is diagonal. Now if one want to get a matrix expression like  $P_{ij}(\boldsymbol{k}) $, one has to go back to the original plane-wave basis $\left \{|i\rangle \right\}$. 

We suppose
\begin{equation}
|\psi(\bk)\rangle=a_{1,k} |e_1(\bk) \rangle +a_{2,k} |e_2(\bk) \rangle
\end{equation}
then
\begin{equation}
\begin{aligned}
&\langle \psi(\bk_1)|\psi(\bk_2)\rangle=\\
&a_{1,k_1}^{*} a_{1,k_2} \langle e_1(\bk_1) |e_1(\bk_2) \rangle +a_{1,k_1}^{*}a_{2,k_2} \langle e_1(\bk_1)|e_2(\bk_2) \rangle \\
&+a_{2,k_1}^{*}a_{1,k_2} \langle e_2(\bk_1)|e_1(\bk_2) \rangle +a_{2,k_1}^{*}a_{2,k_2} \langle e_2(\bk_1)|e_2(\bk_2) \rangle\\
=&\sum_{i,j=1}^{2} a_{i,k_1}^{*} a_{j,k_2} V_{i,j}(\bk_1,\bk_2)
\end{aligned}
\end{equation}
which means:
\begin{equation}
V_{i,j}(\bk_1,\bk_2)=\langle e_i(\bk_1) |e_j(\bk_2) \rangle=\sum_{l}U^{*}_{m_{i},l}\left(\mathbf{k}_{1}\right) U_{l,m_{j}}\left(\mathbf{k}_{2}\right).
\end{equation}  

In addition, due to the existence of band degeneracy, in order to obtain a well-defined Chern number in the calculation of finite-temperature QMC, we added a perturbation term which breaks time-reversal symmetry, and wrote it into the basis of BM model. Then we added it to the calculation of quantum Monte Carlo, similar to kinetic energy term. The perturbation term is:
\begin{equation}
\delta H=\left[\begin{array}{cc}
\delta \sigma_{z} & 0 \\
0 & \delta \sigma_{z}
\end{array}\right]
\end{equation}
where $\delta=0.01$ meV.

After considering the link matrix $V$ and adding the perturbation term, we obtain a gauge-independent algorithm to calculate the Chern number in the interaction model as Eq.~\eqref{eq:eq2} in the main text in the QMC simulations : $C=\sum_{\square} F_{\square}$. As shown in Fig~\ref{fig:fig3}(d), Chern number of TBG for a $N_k=6\times 6$ system at $T=1.25$ meV (below the thermal transition temperature) is $C=1$
. It's consistent with our expectations that at low temperatures the systems become a QAH-TMI with spontaneous time-reversal symmetry breaking.

\subsection{\uppercase\expandafter{\romannumeral2}. Sign Problem}

For the spin and valley polarized cases studied in this work, it has been proved that the weight of configuration is a real number but not necessarily positive~\cite{zhangMomentum2021}. Fortunately, the sign boundary theory~\cite{zhangSign2021,panSign2022} tells us that for such a model in the chiral limit, the average of sign at zero temperature has a polynomial lower bound, which is related to the ground state degeneracy. This means that even if the QMC simulations do not have all the weight positive, one can still solve the problem with polynomial computation complexity even at very low temperatures.

We choose to reweight by considering the modulus of the weight to perform QMC calculation with sign problem:
\begin{equation}
\begin{aligned}
\langle O\rangle&=\frac{\sum_{\mathcal{C}} O(\mathcal{C}_i) W(\mathcal{C}_i)}{\sum_{\mathcal{C}} W(\mathcal{C}_i)}\\
&=\frac{\sum_{\mathcal{C}} O(\mathcal{C}_i) s(\mathcal{C}_i)\,|W(\mathcal{C}_i)| \,/ \sum_{\mathcal{C}}|W(\mathcal{C}_i)|}{\sum_{\mathcal{C}} s(\mathcal{C}_i)|W(\mathcal{C}_i)|\,/ \sum_{\mathcal{C}}\,|W(\mathcal{C}_i)|} \\
&\equiv \frac{\langle O s\rangle^{\prime}}{\langle s\rangle^{\prime}} 
\end{aligned}
\end{equation}
where $W(\mathcal{C}_i)$ is the weight of configuration and $s(\mathcal{C}_i)=\frac{W(\mathcal{C}_i)}{|W(\mathcal{C}_i)|}$. $|W(\mathcal{C}_i)|$ is the new weight after reweighting. The average of sign is defined: $\langle s\rangle^{\prime}=\sum_{\mathcal{C}} s(\mathcal{C}_i)|W(\mathcal{C}_i)|\,/ \sum_{\mathcal{C}}\,|W(\mathcal{C}_i)|$

\begin{figure}
	\centering
	\includegraphics[width=0.99\linewidth]{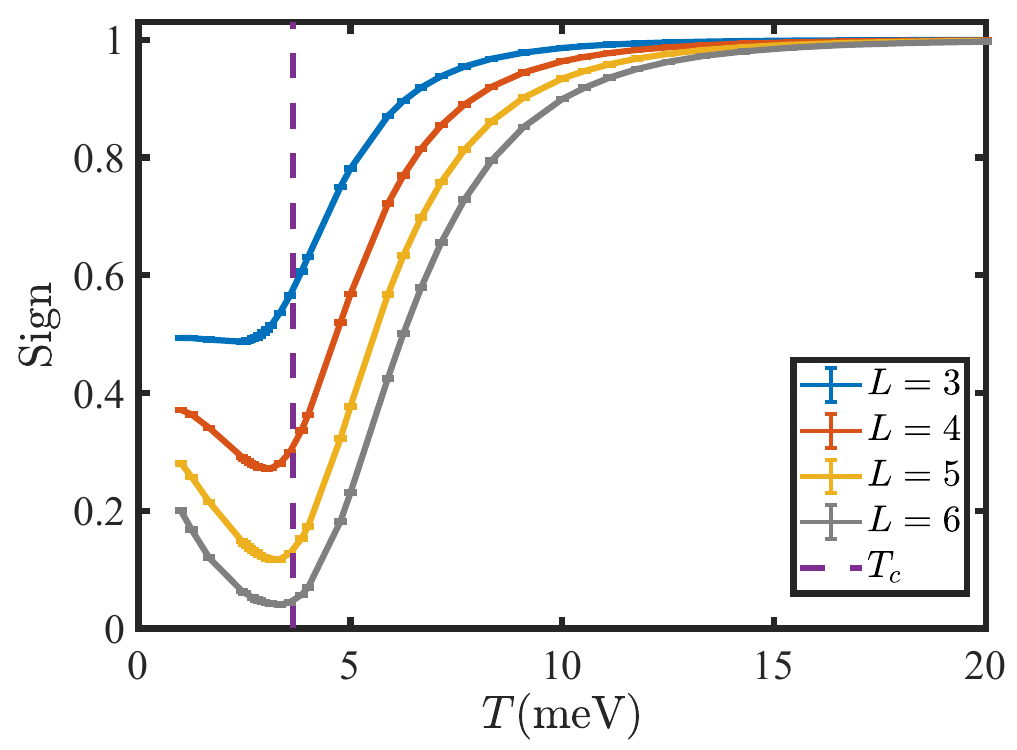}	
	\caption{The average sign of the momentum-space QMC simulation for TBG
	 with $L=3,4,5,6$. As the temperature approaches zero, the average sign approaches a finite value rather than decays exponentially to zero. As shown in Ref.~\cite{zhangSign2021}, it also decays algebraically with respect to the system size $L$.}
	\label{fig:sign}
\end{figure}


As shown in Fig.~\ref{fig:sign}, with decreasing temperature , the average sign does not decay exponentially to zero, but gradually reaches a finite value, where we can perform monte carlo simulation accurately. This is consistent with the conclusion of the sign bound theory~\cite{zhangSign2021,panSign2022}: in the zero temperature limit, the sign has a lower bound related to the ground state degeneracy. In addition, since the ground state  degeneracy of the model we calculated is a polynomial rather than an exponential function of the system size $L$, the average sign does not decay exponentially but algebraically as $L$ increases. As we discussed in Ref~\cite{zhangSign2021}, the peak of average sign here is close to the position of transition temperature $T_c$. Similar behaviors were seen in Ref~\cite{mondainiQuantum2022}. 

\subsection{\uppercase\expandafter{\romannumeral3}. Chern basis}

Since we have $\Sigma_z H(\bk) \Sigma_z =-H(\bk)$, where $\Sigma_z = I_{nG} \otimes I_2 \otimes \tau_z$. Here $I_{nG}$ is a $nG \times nG$  identity matrix of momentum index, where $nG$ is the number of momentum points after truncating the BM model. And $I_2$ is the identity matrix of layer, while $\tau_z$ is the Pauli $z$ matrix of flat band index. Then we can choose chern basis:
\begin{equation}
|u_{\pm}\rangle =|u_1\rangle \pm \Sigma_z |u_1\rangle
\end{equation}
where $|u_1\rangle$ is the eigenstate of $H(\bk)$ at band basis. Now it's clear that  $H|u_{\pm}\rangle=0$ and $\langle u_{+}|u_{-}\rangle=0$. And then we could normalize $|u_{\pm}\rangle$ and we get chern basis $|u_{\pm}\rangle$, where form factor  $\lambda_{m,n}(\mathbf{k}, \mathbf{k}+\mathbf{q}+\mathbf{G})$ is diagonal $2 \times 2$ matrix for each $\bk,\bq,\bG$. Then we can  express it as: $\lambda_{m}(\mathbf{k}, \mathbf{k}+\mathbf{q}+\mathbf{G})$.

If we further fix the gauge of $C_2 T$:
\begin{equation}
|u_1 (\bk)\rangle \rightarrow e^{i \theta(\bk)}|u_1 (\bk)\rangle
\end{equation}
where $\langle u_1(\bk)|O_{C_2 T}|u_1 (\bk)\rangle = e^{-i 2\theta(\bk)}$ and $O_{C_2 T} = I_{nG} \otimes I_2 \otimes \tau_x K$ where $K$ is complex conjugation. So we could just choose $e^{i \theta(\bk)}=(\sqrt{\langle u_1(\bk)|O_{C_2 T}|u_1 (\bk)\rangle})^*$.  Then finally the two diagonal elements of form factor are complex conjugate to each other in chern basis $|u_{\pm}\rangle$. That means: $\lambda_{m}(\mathbf{k}, \mathbf{k}+\mathbf{q}+\mathbf{G})=\lambda_{-m}^{*}(\mathbf{k}, \mathbf{k}+\mathbf{q}+\mathbf{G})$

\begin{figure}[h]
	\centering
	\includegraphics[width=0.99\linewidth]{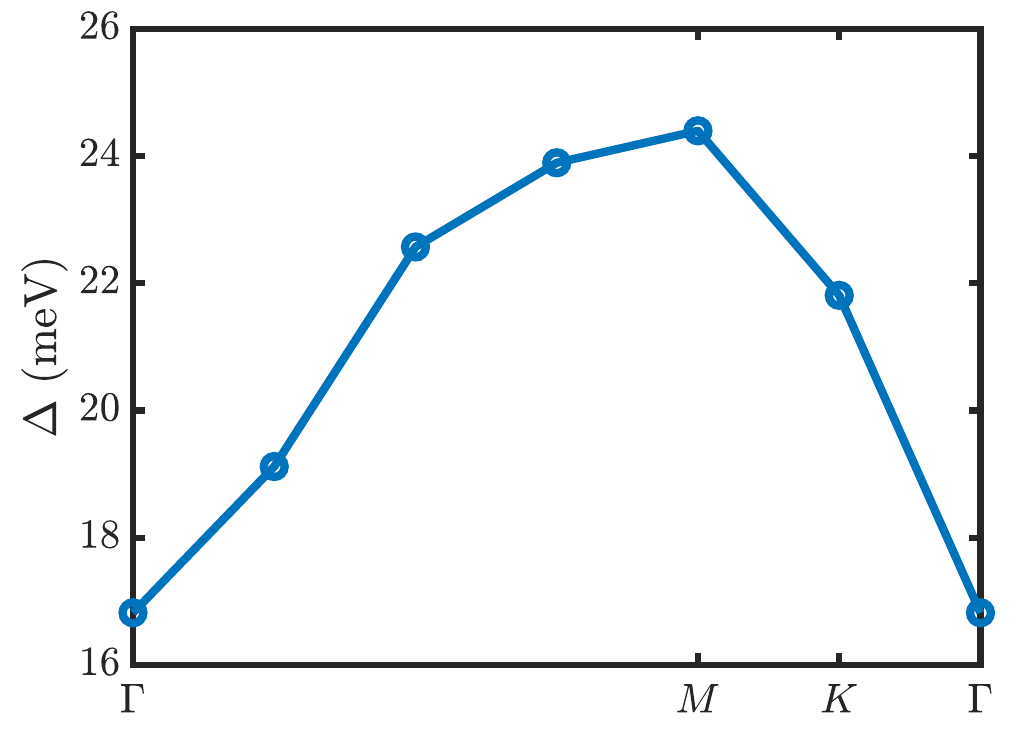}
	\caption{Single-particle excitation gap along the high-symmetry path of TBG  with $L=6$ 
		with $L=6$ from exact solution.}
	\label{fig:singlegap}
\end{figure}
\subsection{\uppercase\expandafter{\romannumeral4}. Single-Particle Excitation Gap}

We obtain the single-particle excitation gap from exact solution by calculating the commutator \cite{bernevig2020tbg5,panDynamical2022}. 

For $\nu=0$, we know that the ground state satisfy :\begin{equation}
	O_{\mathbf{q}+\mathbf{G}} |\Psi\rangle=0
\end{equation}
  
This is also consistent with the results in Fig.~\ref{fig:fig2} and ED result. In the zero temperature limit, the energy of the semidefinite Hamiltonian is zero.

then:
\begin{equation}
	\left[H_{I}, d_{\bkp, n }^{\dagger}\right]|\Psi\rangle=\frac{1}{2 \Omega_{\mathrm{tot}}} \sum_{m_2} R_{m_2 n}^{\eta}(\bkp) d_{\bkp, m_{2} }^{\dagger}|\Psi\rangle
\end{equation}
where
\begin{equation}
\begin{aligned}
	R_{m_1 n_1}^{\eta}(\bk)=\sum_{ m,\bq,\bG,|\bq+\bG|\neq 0}&V(\mathbf{q}+\mathbf{G})\lambda^{*}_{m_1,m  }(\bk,\bk+\bq+\bG)\\ &\lambda_{n_1,m  }(\bk,\bk+\bq+\bG) \
	\end{aligned}
\end{equation}

Diagonalize$\frac{R_{m_1 n_1}^{\eta}(\bk)}{2\Omega}$ and we obtain the single-particle excitations, as plotted in the Fig.~\ref{fig:singlegap}. It can be seen that the temperatures of our TBG calculations are both below the single-particle excitation.

\subsection{\uppercase\expandafter{\romannumeral5}. The implementation of the exact diagonalization}

In the Chern basis, exact diagonalization gives us the single-particle and the exciton gaps for different cluster sizes. Using the diagonal form factors in Chern basis $\lambda_{mn}\sim\lambda_m\delta_{mn}$, the interaction in Eq.\eqref{eq:eq2} can be written as the summation of normal-ordered two-body terms, one-body terms and constant terms:
\begin{eqnarray}
	H&=&\sum_{\mathbf q,\mathbf G,\mathbf k,\mathbf k',m,m'}\frac{V(\mathbf q+\mathbf G)}{2\Omega}\lambda_m(\mathbf k,\mathbf k+\mathbf q+\mathbf G)\lambda_{m'}(\mathbf k',\mathbf k'-\mathbf q-\mathbf G)\nonumber\\
	&&\qquad \qquad d^\dagger_{\mathbf km}d^\dagger_{\mathbf k'm'}d_{\mathbf k'-\mathbf q,m'}d_{\mathbf k+\mathbf q,m} \nonumber\\
	&&+\sum_{\mathbf q,\mathbf G,\mathbf k,m}\frac{V(\mathbf q+\mathbf G)}{2\Omega}|\lambda_{m}(\mathbf k,\mathbf k+\mathbf q+\mathbf G)|^2 d^\dagger_{\mathbf km}d_{\mathbf km}\nonumber\\
	&&-\sum_{\mathbf k,\mathbf k',\mathbf G,m,m'}\frac{V(\mathbf G)}{2\Omega}\lambda_{m}(\mathbf k,\mathbf k+\mathbf G)\lambda_{m'}(\mathbf k',\mathbf k'-\mathbf G)d^\dagger_{\mathbf km}d_{\mathbf km} \nonumber\\
	&&+\frac{1}{4}\sum_{\mathbf k,\mathbf k',\mathbf G,m,m'}\frac{V(\mathbf G)}{2\Omega}\lambda_{m}(\mathbf k,\mathbf k+G)\lambda_{m'}(\mathbf k',\mathbf k'-\mathbf G)
\label{EDexpand}
\end{eqnarray}
To perform exact diagonalization, we can use an integer $i\in [0,2L^2-1]$ to label each single particle state $d^\dagger_i\equiv d^\dagger_{\mathbf km}$ whose Chern band index $m$ and momentum $\mathbf k$ are determined by $i$. We make $\mathbf k$ inside the moir\'e Brillouin zone. Eq.\eqref{EDexpand} can be written as
\begin{equation}
    H=\sum_{i,i',j',j} V_2(i,i',j',j) d_i^\dagger d_{i'}^\dagger d_{j'}d_j+\sum_{i}V_1(i)d_i^\dagger  d_i+V_0
    \label{EDsim}
\end{equation}

\begin{figure}[t]
	\centering
	\includegraphics[width=0.9\linewidth]{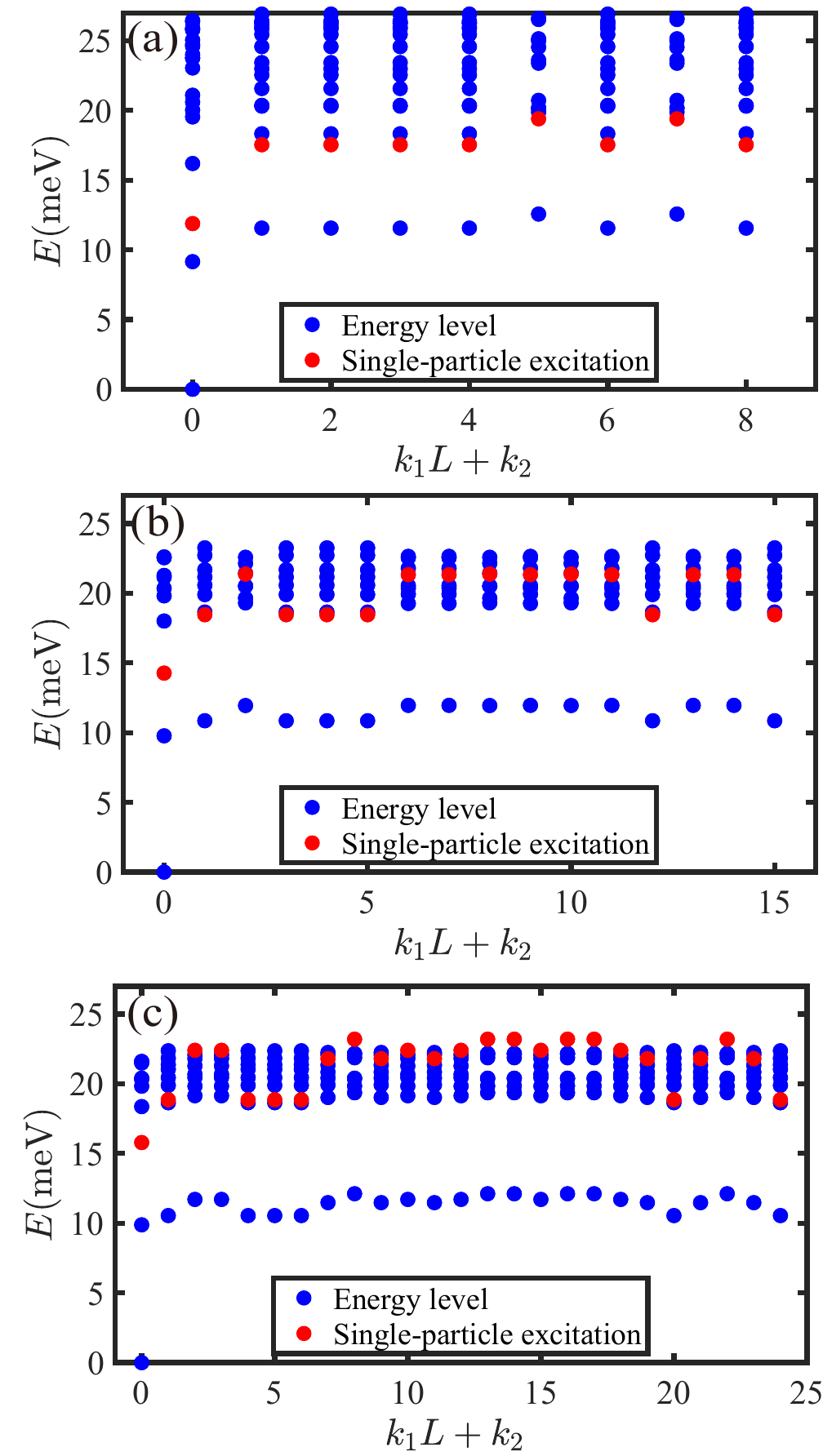}
	\caption{ Single-particle excitation gap and the exciton gaps of TBG with (a) $L=3$, (b) $L=4$, (c) $L=5$ from exact diagonalization method.}
	\label{fig:ed}
\end{figure}

The matrix element between arbitrary manybody states can be obtained from Eq.\eqref{EDsim}. Let $\ket{\psi}$ be a simple direct product state $\ket{\psi}=C^\dagger_{i_{N-1}}...C^\dagger_{i_0}\ket{0}$ whose creation operators are ordered by $i_{N-1}>...>i_1>i_0$. Then $\ket{\psi}$ can also be represented by a string of bits with length $2L^2$ such as $\ket{1101001...}$. Suppose the Hamiltonian has nonzero matrix element between distinct manybody states $\ket{\psi}$ and $\ket{\psi'}$, i.e., there exist $i>j,i'>j'$ such that $\bra{\psi'}d_{j'}^\dagger d^\dagger_{i'}d_i d_j\ket{\psi}\ne 0$, then the matrix element is given by:
\begin{equation}
\begin{aligned}
\bra{\psi'} \hat V_2\ket{\psi}=(-1)^{\mu(i,j)}(-1)^{\mu'(i',j')}&\left[V_2(j',i',i,j)-V_2(j',i',j,i)\right.\\
-&\left.V_2(i',j',i,j)+V_2(i',j',j,i)\right]
\label{ED1}
\end{aligned}
\end{equation}
Where $\hat V_2$ is the two-body term in Eq.\eqref{EDsim}, $\mu(i,j)$ is the number of "1" between the $i$-th and $j$-th bits in the bit-representation of $\ket{\psi}$. This sign factor comes from the Fermi anti-commutation of creation operators. If $\ket{\psi}=\ket{\psi'}$, then $i=i'$ and $j=j'$, and the matrix element involves a sum over all combinations of $i,j$ that do not annihilate the state:
\begin{equation}
\bra{\psi}\hat V_2\ket{\psi}=\sum_{i>j,\psi_i=\psi_j="1"}(V(j,i,i,j)-V(j,i,j,i)-V(i,j,i,j)+V(i,j,j,i))
\end{equation}
Where $\psi_i$ is the $i$-th bit in the bit-representation of $\ket{\psi}$ that can take either "1" or "0". The one-body term $\hat V_1$ in Eq.\eqref{EDsim} only contribute to the diagonal elements of the many-body Hamiltonian matrix:
\begin{equation}
    \bra{\psi}\hat V_1\ket{\psi}=\sum_{i,\psi_i="1"}V_1(i)
    \label{ED3}
\end{equation}
The many-body Hamiltonian matrix can be constructed via Eqs.\eqref{ED1}-\eqref{ED3}. Because the interaction conserves total momentum $\mathbf k$ and the Chern polarization $P_C$, the many-body matrix is block-diagonal with each block composed of many-body states with the same total momentum and $P_C$. By diagonalizing these blocks we can resolve the eigenstates by $\mathbf k$ and $P_C$. 

The spectrum for TBG systems with $L=3,4,5$ are shown in Fig.\ref{fig:ed}. The blue dots represent the spectrum of a system with $L^2$ particles, and the red dots represent the single-particle excitation obtained by the lowest energy states of a system with $L^2+1$ particles. 

\end{document}